\documentclass[aps,twocolumn,showpacs,groupedaddress,nofootinbib]{revtex4}

\usepackage{graphicx}

\begin{document}

%
%
\preprint{MI-15-01}

\title{Enhancement of Br($B_d \to \mu^+\mu^-$)/Br($B_s \to \mu^+\mu^-$)
 in Supersymmetric Unified Models}


\author{Bhaskar Dutta$^1$}
\author{Yukihiro Mimura$^2$}
\affiliation{
$^1$Department of Physics and Astronomy, Mitchell Institute for Fundamental Physics and Astronomy, Texas A\&M University,
College Station, TX 77843-4242, USA \\
$^2$Department of Physics, National Taiwan University,
Taipei 10617, Taiwan
}



\begin{abstract}
We explain the 2.3$\sigma$ deviation in the recent measurements of
the neutral $B$ mesons  decay into muon pairs
from the standard model prediction
in the framework of supersymmetric grand unified models
using  anti-symmetric coupling as a new source of flavor violation.
We show a correlation between the $B_d \to \mu^+\mu^-$ decay and the CP phase in the $B_d\to J/\psi K$ decay
and that their deviations from the standard model predictions can be explained
after satisfying  constraints arising from various hadronic
and leptonic rare decay processes, $B$-$\bar{B}$,  $K$-$\bar{K}$ oscillations data and
electric dipole moments of electron and neutron.
The allowed parameter space
is typically represented by pseudoscalar Higgs mass $m_A\leq 1$ TeV and $\tan\beta_H(\equiv {v_u}/{v_d}) \alt 20$
for squark and gluino masses around 2 TeV.
\end{abstract}



\maketitle


\section{Introduction}

The recent measurements of the branching fractions of the rare $B$ meson decays,
$B_s^0 \to \mu^+\mu^-$ and $B_d^0 \to \mu^+\mu^-$
showcase impressive achievements of the LHC experiments \cite{CMS:2014xfa}.
The ratios of the experimental measurements and the standard model (SM) predictions \cite{Bobeth:2013uxa} are
\begin{eqnarray}
{{\rm Br}(B_s \to \mu^+\mu^-)_{\rm exp/SM}} &=& 0.76^{+0.20}_{-0.18},
\\
{{\rm Br}(B_d \to \mu^+\mu^-)_{\rm exp/SM}} &=& 3.7^{+1.6}_{-1.4}.
\end{eqnarray}
Both measurements are consistent with the SM predictions within the errors,
though Br($B_d \to \mu^+\mu^-$) seems to be a bit larger than the SM prediction.

The ratio of the fractions,
\begin{equation}
R \equiv \frac{{\rm Br}(B_d \to \mu^+\mu^-)}{{\rm Br}(B_s \to \mu^+\mu^-)},
\end{equation}
has less theoretical errors in the SM compared to each fraction
because the ratio of the decay constants has less ambiguity compared to each decay constant
in the lattice calculation.
The prediction for the SM (and for the models with minimal flavor violation \cite{Buras:2000dm}) is
\begin{equation}
R_{\rm SM} \simeq \frac{\tau_{B_d}}{\tau_{B_s}}\frac{M_{B_d}}{M_{B_s}}
\frac{f_{B_d}^2}{f_{B_s}^2}\left|\frac{V_{td}}{V_{ts}}\right|^2,
\end{equation}
where $\tau_B$, $M_B$, and $f_B$ are the lifetime, mass, and decay constant
of the respective mesons,
 and \cite{CMS:2014xfa,Bobeth:2013uxa}
\begin{equation}
R_{\rm SM} = 0.0295^{+0.0028}_{-0.0025}.
\end{equation}
The ratio using the experimental measurements
is
\begin{equation}
R_{\rm exp} = 0.14^{+0.08}_{-0.06},
\end{equation}
and the experimental result shows deviation from the SM prediction
at 2.3$\sigma$.

These rare decays are induced radiatively in the SM,
and thus, they are sensitive to the new physics, and their measurements provide us a direction
in which the SM can be extended to the models beyond SM \cite{Hou:2013btm}.
The deviation from the SM prediction is not very significant statistically at present, however,
it is meaningful to investigate the models which can enhance the ratio $R$
since  the usual source of  flavor chaining neutral currents (FCNC) does not produce any enhancement naturally.
In this paper,
we suggest a possible source to explain the enhancement of the ratio naturally
in the supersymmetric (SUSY) standard model,
and investigate the implications of the models in the possible unified frameworks,
such as SU(5) and SO(10) grand unified models.

In order to modify the ratio $R$ compared to the SM prediction,
one needs a new type of FCNC source which is different from Cabibbo-Kobayashi-Maskawa (CKM) flavor mixings.
The main concerns regarding the  enhancement of $R$ are the followings:

\begin{enumerate}
\item
How natural is that to have a larger $b \to d$ transition compared to the $b\to s$ transition in the presence of a new FCNC source to enhance $R$?
In fact, the $B_s$-$\bar B_s$ mixings and the $B_s \to J/\psi\phi$ decay are consistent with the SM predictions \cite{Bs-phase}
and it seems that there is not a large source of FCNC in the $b\to s$ transition.

\item
The mass difference of $B_d$ and the CP violation in the $B_d$-$\bar B_d$ mixings from the
the experimental measurements of the $B_d$-$\bar B_d$ oscillations and the $B_d \to J/\psi K$ decay
are consistent with the SM prediction.
How can  large modifications of them be avoided if the Br($B_d \to \mu^+\mu^-$) is enhanced by a new FCNC source?

The SM prediction of $\sin 2\beta$ has a slight difference from the experimental measurements\footnote{
Recently, LHCb released their new analysis of the CP violation in the $B_d \to J/\psi K$ decay \cite{Aaij:2015vza}:
\begin{equation}
\sin2\beta = 0.746 \pm 0.030.
\end{equation}
The world average of the CP phase becomes larger by the new LHCb result,
and the deviation from the SM prediction becomes less, though it is still a smaller value than it.
}
from the $B_d \to J/\psi K$ decay \cite{CKMfit}:
\begin{eqnarray}
\sin2\beta&=& 0.692^{+0.020}_{-0.018} \ ({\rm BaBar\ \&\ Belle\ exp.}),
\label{sin2b-exp}
\\
\sin2\beta &=& 0.774^{+0.017}_{-0.036} \ ({\rm SM \ prediction}).
\label{sin2b-SM}
\end{eqnarray}
Can the slight difference be consistent with a modification of $R$?

\item
The experimental results of the $b\to s\gamma$ and $b\to d \gamma$ decays are consistent with the SM prediction.
The ratio of Br($b\to d \gamma$)/Br($b\to s\gamma$) in SM is also related to $|V_{td}/V_{ts}|^2$
up to hadronic uncertainty \cite{Ali:1998rr}.
The ratio, by using the experimental measurements \cite{Wang:2011sn,Agashe:2014kda},
is
Br($b\to d \gamma$)/Br($b\to s\gamma$) = $0.040\pm 0.009\pm 0.010$.
How natural can the enhancement of $B_d \to \mu^+\mu^-$ be without enhancing $b\to d \gamma$?

\end{enumerate}

\section{FCNC induced by anti-symmetric couplings}

In SUSY models, too much FCNCs are generated in general,
and thus, the flavor universality of the SUSY breaking mass parameters
are often assumed.
In such a framework,
the renormalization group evolution can generate off-diagonal elements in the sfermion mass matrices
and  FCNCs are induced.
If the CKM quark mixing is the only source to generate the off-diagonal elements,
$R$ is not modified even though the individual branching fractions can be modified.
In unified models, new particles can propagate in the loops and it can generate
a new type of flavor violation source \cite{Borzumati:1986qx,Dutta:2006zt,Dutta:2007ai}.
In simple models, this new FCNC source can induce $b$-$s$ transitions,
and thus, the ratio $R$ is rather reduced. 
In this paper, we suggest a model to provide a possible explanation for enhancing $R$.

In general, the Yukawa coupling $Y_{ij}$ to the quarks can induce off-diagonal elements in the squarks mass matrices
by RGE in the form of
\begin{equation}
(M_{\tilde q}^2){}_{i\neq j} = - \frac{C}{8\pi^2} Y_{ik} Y^*_{jk} (3 m_0^2 + A_0^2) \ln \frac{M_*}{M_X},
\end{equation}
where $m_0$
is a universal scalar mass, $A_0$ is a universal scalar trilinear coupling,
$M_*$ is a cutoff scale, $M_X$ is the mass of a heavy field which propagate in the loop,
and $C$ is a group weight factor.
It is important to note that
the gluino and squark masses should be heavy due to the LHC results,
and thus, the amount of the induced FCNC becomes less.
The discovery of the 125 GeV Higgs boson also pushes up the squark masses.
If the mass of the squarks and gluino are O(10) TeV, it is hard to extract the off-diagonal elements from the flavor data.
However, if the squark and gluino masses are about 2 TeV, the scalar trilinear coupling $A_0$ has to be large ($\sim$ 5 TeV)
to obtain the Higgs mass to be 125 GeV,
and then, the off-diagonal elements are generated (even if $m_0$ is small)
and the FCNCs are induced slightly.
Therefore, the SUSY contribution can be consistent with the experimental results of many of the FCNC processes,
but a slight excess can be observed in a process whose amplitude can have a enhancement factor.
We remark that the circumstances are changed from the literatures a few years ago.

The RGE-induced off-diagonal elements are characterized by $YY^\dagger$,
and it can be parameterized as
\begin{equation}
Y Y^\dagger \propto U {\rm diag.} (k_1, k_2, 1) U^\dagger,
\end{equation}
where $U$ is a diagonalizing unitary matrix of $Y$,
and $k_1$, $k_2$ are the ratios of eigenvalues of $Y Y^\dagger$.
Using usual mixing parameterization in $U$, the off-diagonal elements can be expressed as
\begin{eqnarray}
\!\!M_{23}^2 &\!\!\propto&\!\! -\frac12 \sin2\theta_{23} , \\
\!\!M_{13}^2 &\!\!\propto&\!\! -\frac12 k_2 \sin2\theta_{12}\sin\theta_{23} + e^{i\delta} \sin\theta_{13}\cos\theta_{23} , \\
\!\!M_{12}^2 &\!\!\propto&\!\! -\frac12 k_2 \sin2\theta_{12}\cos\theta_{23} - e^{i\delta} \sin\theta_{13}\sin\theta_{23},
\end{eqnarray}
where the hierarchy of eigenvalues $k_1 \ll k_2 \ll 1$ is assumed.
The popular source of the flavor violation is the Dirac neutrino Yukawa coupling matrix $Y_\nu$,
which can contain the large mixing angles realigning to neutrino mixings.
In SU(5) GUT models, $Y_\nu$ can also induce the off-diagonal elements in the right-handed
down-type squark mass matrix via colored Higgs loop diagram.
Thus, in the popular scenario, the $b$-$s$ FCNC is induced relating to the large atmospheric neutrino mixing.
However, it turns out that via the measurements of $B_s$-$\bar B_s$ oscillations
the new FCNC contribution in the mixing amplitude should be small \cite{Bs-phase}.
In addition to that, the 12 and 13 elements are naively same order if the mixing angles in $U$ are related to the neutrino mixings, and thus
the large $b$-$d$ FCNC needs a cancellation between two terms in $M_{12}$ due to the experimental results of $K$-$\bar K$ mixing.
Surely, the mixing angles in $Y_\nu$ is not directly same as the neutrino mixing since the neutrino mass matrix is
proportional to $Y_\nu M_R^{-1} Y_\nu^\dagger$, where $M_R$ is the right-handed neutrino mass matrix.
Therefore, there can be a solution to enlarge the ratio $R$ for the $B_{d,s} \to \mu^+\mu^-$ decays,
but one should admit that it is not a natural solution in this popular source of the Dirac neutrino coupling.
We thus suggest a new source of flavor violation to enlarge $R$.

We now consider the following anti-symmetric coupling matrix $h^\prime$ under the flavor indices for the left-handed quark doublet $q$:
\begin{equation}
h_{ij}^\prime q_i q_j ({\bf 3},{\bf 3}, -\frac13),
\quad {\rm or}\quad
h_{ij}^\prime q_i q_j (\overline{\bf 6},{\bf 1}, -\frac13).
\end{equation}
Here, $({\bf 3},{\bf 3}, -\frac13)$ and $(\overline{\bf 6},{\bf 1}, -\frac13)$ are new fields
whose representations are denoted under the SM gauge group, $SU(3)_c\times SU(2)_L \times U(1)_Y$.
They can arise in grand unified models, SU(5), SO(10), and so on, as we will study later.
Denoting
\begin{equation}
h^\prime = \left(
 \begin{array}{ccc}
  0 & a & -b \\
  -a & 0 & c \\
  b & -c & 0
 \end{array}
\right),
\end{equation}
one obtains
\begin{equation}
h^\prime h^{\prime\dagger} = \left(
 \begin{array}{ccc}
  |a|^2+|b|^2 & -bc^* & -ac^* \\
  -b^*c & |a|^2+|c|^2 & -ab^* \\
  -a^*c& -a^*b & |b|^2+|c|^2
 \end{array}
\right).
\end{equation}
We then find interesting features in the off-diagonal elements arising from the anti-symmetric coupling:

\begin{enumerate}
\item
In the case of a naive hierarchy $|b| < |c|$ in the $h^\prime$ coupling,
one obtains an inverted hierarchy in the off-diagonal elements using the RGEs since
 $|(h^\prime h^{\prime\dagger})_{13}| > |(h^\prime h^{\prime\dagger})_{23}|$,
 and thus it can be expected that the $b$-$d$ FCNC is larger than the $b$-$s$ FCNC.

\item
The magnitudes of two out of three off-diagonal elements ($12$, $13$ and $23$) in $h^\prime h^{\prime\dagger}$ are correlated.
 For example, if $12$ element is zero, one of 13 and 23 elements of $h^\prime h^{\prime\dagger}$ is zero.
 One can easily enhance $b\to d$ transition (but not $b\to s$) after satisfying $K$-$\bar K$ data.

\end{enumerate}

These two features nicely explain how $R$ is enhanced naturally using the RGE-induced FCNC.
In order to illustrate these features, we plot  $B_d \to \mu^+\mu^-$ vs $B_s \to \mu^+\mu^-$
by imposing the anti-symmetric coupling $h^\prime$ (Fig.1).
We use  universal scalar masses for squark and slepton fields, $m_{0}=2$ TeV and
the unified gaugino mass, $m_{1/2} = 1$ TeV.
The universal scalar trilinear coupling $A_0$ is chosen to make the Higgs mass
to be 125 GeV ($A_0 \simeq$ 5 TeV).
However, we use non-universal SUSY breaking Higgs masses,
and we choose the Higgsino mass $\mu = 3$ TeV and the CP odd Higgs mass $m_A = 1$ TeV. The values of $a$, $b$ and $c$ are $<$ 1.
%
%
The ratio of the Higgs vacuum expectation values, $\tan\beta_H $ is chosen
not to be very large (here we use $\tan\beta_H = 20$) to make it consistent with the experimental measurement.
We will explain the reason for these choices later.
In this plot, the naive hierarchies among $a$, $b$ and $c$ (such as $|a|, |b|<|c|$) is not assumed
to illustrate the second feature.
As it is expected, the green points (which correspond to the choice of small 12 off-diagonal element) appear like a cross or $\dagger$ symbol.
We note that even in the case where there is no new 23 FCNC source, $B_s \to \mu^+\mu^-$ is enhanced due to the
chargino contribution.

\begin{figure}[t]
\begin{center}
\includegraphics[width=.48\textwidth,clip]{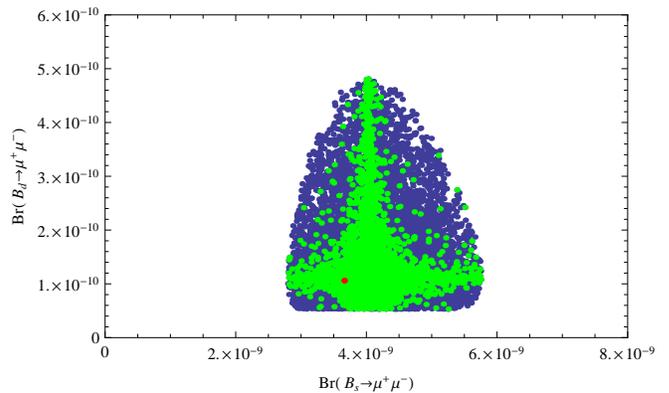}
\caption{$B_s \to \mu^+\mu^-$ and $B_d \to \mu^+\mu^-$
induced by the anti-symmetric FCNC source are shown.
The setup is detailed in the text.
The blue points correspond to the case of randomly generated $a,b,c$ (with phases)
using the anti-symmetric coupling,
and the green points  satisfy $K$-$\bar K$ mixing data
($\Delta M_K$ and $\epsilon_K$).
%
The red point shows the SM prediction.
}
\end{center}
\end{figure}

\subsection{$B_d$-$\bar B_d$ mixings}

\begin{figure}[t]
\begin{center}
\includegraphics[width=.43\textwidth,clip]{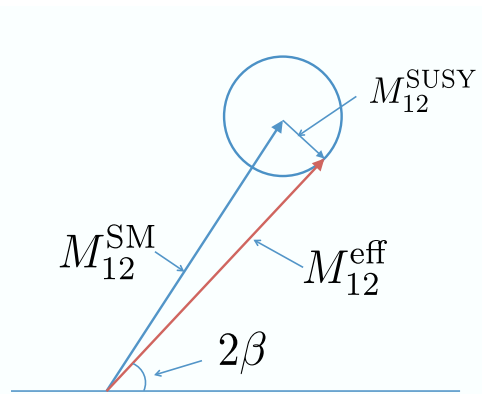}
\caption{
The illustration of the SM and SUSY contribution of the mixing amplitude
of $B$-$\bar B$ mixing in the complex plane.
The detail is explained in the text.
}
\end{center}
\end{figure}

The $B_d$-$\bar B_d$ mixing amplitude
is given as
\begin{equation}
M_{12}^{\rm eff} = M_{12}^{\rm SM} + M_{12}^{\rm SUSY},
\end{equation}
and the mass difference of the $B_d$ meson and
a CP phase of $B_d \to J/\psi K$ decay are
obtained
as $\Delta M = 2|M_{12}^{\rm eff}|$, and $2\beta = {\rm arg} M_{12}^{\rm eff}$.
We show a geometrical illustration in the complex plane of the mixing amplitude $M_{12}$ in Fig.2.
The magnitude of $M_{12}^{\rm SUSY}$ can be calculated when the SUSY particle spectrum is fixed,
while the phase of $M_{12}^{\rm SUSY}$ is free depending on the phase parameter of the new source for FCNC.
Therefore, the arrowhead of $M_{12}^{\rm eff}$ tracks the circle shown in the figure.
We comment that ${\rm arg}\, M_{12}^{\rm SUSY} =  {\rm arg}\, M_{12}^{\rm SM}$ is satisfied in the case of minimal flavor violation,
and thus the CP phase of the mixing amplitude is not changed from SM prediction.

To satisfy the experimental results of the $B_d$-$\bar B_d$ oscillation,
$|M_{12}^{\rm SUSY}|$ has to be sufficiently small,
but a contribution of small size can be expected due to a
slight discrepancy of the $\sin2\beta$ measurement in Eqs.(\ref{sin2b-exp}),(\ref{sin2b-SM}).
We remark that the modification of the magnitude of $M_{12}^{\rm eff}$ (i.e., the  mass difference of $B$-$\bar B$ mesons)
is expected to be rather small in the direction of the phase modification by $M_{12}^{\rm SUSY}$
as illustrated in Fig.2.
On the other hand, to enhance the ratio $R$,
a sizable SUSY contribution to the $B_d \to \mu^+\mu^-$ decay amplitude.
We need to explain how such a situation is reproduced by the left-handed quark FCNC source.
We note that the SUSY contribution of the box diagram for the $B$-$\bar B$ mixing amplitude is now suppressed
(roughly 10\% of SM contribution)
due to the LHC constraint of qluino and squarks.
The key is thus how the SUSY contribution of the $B \to \mu^+\mu^-$ decay is enhanced
even when the SUSY particles are heavy $\sim 2$ TeV.
The SUSY contributions to the $B \to \mu^+\mu^-$ amplitudes are dominated by the
Higgs-penguin diagram \cite{Choudhury:1998ze}.
The Higgs FCNC coupling via the non-holomorphic finite correction terms due to SUSY breaking are
enhanced for a large $\tan\beta_H$ since the mass eigenstates of the down-type quarks are modified
by the non-holomorphic correction terms.
As a result, the Higgs-penguin contribution can be the same order of the SM contribution
even with the heavy gluino and squarks masses.

The Higgs FCNC coupling can also contribute to the $B$-$\bar B$ mixing amplitude
via the double Higgs-penguin mediated diagram \cite{Hamzaoui:1998nu}.
In fact, the SUSY contribution $M_{12}^{\rm SUSY}$ is enhanced if
there are FCNC sources in both left- and right-handed quarks.
However, if it is in only one of left- and right-handed sector,
the mixing amplitude is not enhanced and the box contribution becomes dominant.
On the other hand,
the amplitudes of $B\to \mu^+\mu^-$ decay can be enhanced (compared to the SM amplitude) even if there is only left-handed
quark FCNCs arising due to the off-diagonal elements in the squark mass matrices.
%
%
Since there are FCNCs originated from CKM mixings in the left-handed squarks sector,
the off-diagonal elements in the right-handed down-type squarks should not be there
in order to naturally satisfy the experimental results for the $B_d$-$\bar B_d$ mixings and the phase in the $B_d \to J/\psi K$ decay
if there is a sizable $b$-$d$ transition to enhance the amplitude of $B_d \to \mu^+\mu^-$.
If this is the case, the SUSY contribution to the $B_d$-$\bar B_d$ mixing amplitude
is dominated by a box diagram,
and $|M_{12}^{\rm SUSY}|/|M_{12}^{\rm SM}|$ is roughly 10\% for squark and gluino masses  around 2 TeV,
which can explain the slight modification of the CP phase of $B_d \to J/\psi K$ decay.

The flavor violating mass (i.e., off-diagonal element of squark mass matrix) is inserted twice
in the $B$-$\bar B$ mixing amplitude from the box diagram,
while it is inserted once in the Higgs-penguin diagram for the $B \to \mu^+\mu^-$.
Therefore, the phases of the SUSY contributions to those two amplitudes are different but related.
The experimental result of $B_s \to \mu^+\mu^-$ branching fraction is consistent with the SM prediction.
Therefore, we choose the SUSY parameters to make the SUSY contribution to $B_s \to \mu^+\mu^-$
less if the new FCNC source is absent (namely, at the center of the cross in Fig.1).
More concretely, for example, we choose $\tan\beta_H$ to be a value for benchmark SUSY mass parameters
($m_A=1$ TeV, $m_0=2$ TeV, $m_{1/2} = 1$ TeV)
in order to make $B_s \to \mu^+\mu^-$ consistent with the experiment.
Then, the SUSY contribution to $B_d \to \mu^+\mu^-$ amplitude in the case of no new FCNC source is also small
because the ratio of the branching fraction is fixed for the minimal flavor violation.
In that case, the enlarged SUSY contribution of $B_d \to \mu^+\mu^-$ amplitude from the flavor violation is almost directly related to
the phase in the FCNC source.
For making one round of the circle in the $M_{12}^{\rm eff}$ plane in Fig.2,
the phase of the SUSY contribution of $B_d\to\mu^+\mu^-$ is changed a half round.
As a result, a typical correlation between $B_d\to \mu^+\mu^-$
and $\sin2\beta$ modifications is obtained as shown in Fig.3.
The Fig.3 is drawn by using the same SUSY parameters before.
Interestingly, for smaller values of the effective $\sin2\beta$ (which is consistent with the measured value),
 two distinctive regions of $B_d\to\mu^+\mu^-$ appear,
 and we find that $\sin2\beta$ is decreased for the most enhanced values of Br($B_d\to \mu^+\mu^-$).
We note that such qualitative behavior (namely, the ``shape" of the plots in Fig.3) is not sensitive to
the SUSY mass parameters under the above setup to make Br($B_s \to \mu^+\mu^-$) consistent with the experiment
(though the shape can be collapsed depending on the size of the anti-symmetric FCNC source)
as far as the size of $M_{12}^{\rm SUSY}/M_{12}^{\rm SM}$ is up to 10\%
to satisfy the mass differences of $B$-$\bar B$ mesons.
This can be easily understood by the simple geometrical explanation above
and there is only one phase in the new left-handed $b$-$d$ FCNC.

\begin{figure}[t]
\begin{center}
\includegraphics[width=.48\textwidth,clip]{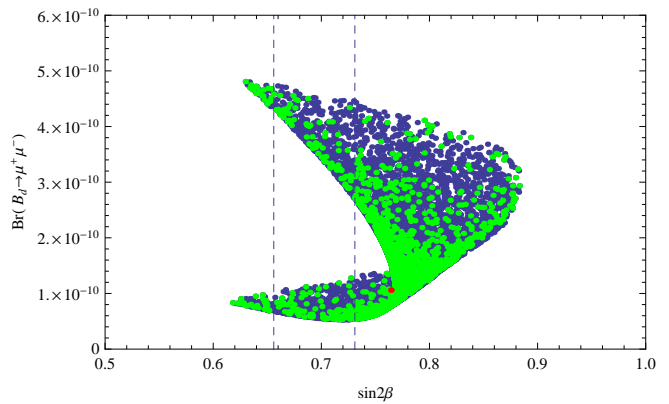}
\caption{
We show the correlation between $B_d \to \mu^+\mu^-$ and $B_d$-$\bar B_d$ mixing
for the choice of SUSY parameters described in the text.
The blue plots are for the randomly generated $a,b,c$ in the anti-symmetric coupling,
and the green points satisfy all the experimental data, such as $K$-$\bar K$ mixing, $B_s$-$\bar B_s$ mixing,
and $b\to d\gamma$, $b\to s\gamma$.
The red point shows the SM prediction.
 The dashed lines  show the 2$\sigma$ region of the experimental measurement of $\sin2\beta$.}
\end{center}
\end{figure}

\subsection{Constraints from $b\to d\gamma/b\to s\gamma$}

The $b\to d \gamma$ and $b\to s \gamma$ processes are also important for constraining the
flavor violations beyond the standard model.
Even in the minimal flavor violation, each branching fraction can be modified compared to the SM prediction,
but the ratio of the fractions is  same as the SM.
If the ratio of the branching fractions
of $B_d \to \mu^+\mu^-$ and  $B_s \to \mu^+\mu^-$
is modified,
the ratio of the branching fractions of $b\to d \gamma$ and $b\to s \gamma$
is also modified in principle.
The experimental measurements of those branching fractions are consistent with the SM prediction
within the experimental errors.
Since the CP phases of the flavor violating decays are not yet observed,
we can have solutions for the partial decay widths to satisfy the experimental results,
even though there are sizable contribution to the amplitudes from new physics.
However, if the experimental results are naturally satisfied,
the new physics contribution should be small.
The amplitude for the $b \to q \gamma$ ($q=d,s$) process
is roughly proportional to $\tan\beta_H$.
Therefore, if $\tan\beta_H$ is not as large as 30-50,
the SUSY contribution can be small even if the squark masses
are then at the edge of the current LHC bounds in the minimal supergravity scenario.
The amplitudes for the $B_q \to \mu^+\mu^-$ process
is proportional to $\tan^3\beta_H/m_A^2$,
and thus,
the SUSY contribution can be comparable to the SM prediction for smaller CP odd Higgs mass $m_A < 1$ TeV
even for small $\tan\beta_H < 20$. The  search prospect for a CP odd Higgs boson is given in
\cite{CMS:2014yra, Aad:2015wra}.

We note that for the minimal supergravity SUSY breaking boundary conditions,
where the SUSY breaking Higgs masses are same as the other SUSY breaking scalar masses at the grand unified scale,
$m_A$ is roughly same as the  squark masses (except for large $\tan\beta_H\sim 50$).
Therefore, in order to obtain large $R$ with natural fits for $b\to d\gamma/b\to s\gamma$, we need the non-universal Higgs scalar mass scenario
for the SUSY breaking.

\section{Unified Models and implications}

The fields
$({\bf 3},{\bf 3}, -\frac13)$,
$(\overline{\bf 6},{\bf 1}, -\frac13)$
which provide the antisymmetric couplings to the left-handed quark doublets
can be unified with the Higgs representations in
unified theories, such as SU(5) and SO(10).
In SU(5),
\begin{equation}
{\bf 10} \times {\bf 10} = \bar{\bf 5}_s + \overline{\bf 45}_a + \overline{\bf 50}_s,
\end{equation}
and
in SO(10),
\begin{equation}
{\bf 16} \times {\bf 16} = {\bf 10}_s + {\bf 120}_a + {\bf 126}_s,
\end{equation}
where $s$ and $a$ stand for symmetric and anti-symmetric, respectively.
Therefore, the Yukawa couplings with
${\bf 45}$ in SU(5) and ${\bf 120}$ in SO(10)
can provide the anti-symmetric sources for the FCNCs.

\begin{table}
\center
\begin{ruledtabular}
 \begin{tabular}{|c|c|c|c|c|c|}
  \hline
  && $c.c.$ & SU(5) & SU(5)$\times $U(1)$_X$ & SO(10)\\ \hline
  $(qq)_a$ & $({\bf 3},{\bf 3}, -\frac13)  $ & $q \ell$  &$  {\bf 45}$ & $( {\bf 45},-2)$& ${\bf 120}$\\
  $(qq)_a$ & $(\overline{\bf 6},{\bf 1}, -\frac13) $ & $u^c d^c$  & $ {\bf 45}$ & $( {\bf 45},-2)$&${\bf 120}$\\
  $q u^c$ & $({\bf 8},{\bf 2}, \frac12)$ & $q d^c$   & $ {\bf 45}$ &$(\overline{\bf 5},2)$&${\bf 120}$\\
  $q \nu^c$ & $(\overline{\bf 3},{\bf 2}, -\frac16)$ & $\ell d^c$   & $\overline{\bf10}$&$( {\bf 45},-2)$&${\bf 120}$\\
  $q e^c$ & $(\overline{\bf 3},{\bf 2}, -\frac76)$ & $\ell u^c$  & $ {\bf45}$&$(\overline{\bf 10},-6)$&${\bf 120}$\\
 \hline
 \end{tabular}
 \end{ruledtabular}
 \caption{Five candidates for the anti-symmetric bi-fermion coupling to enhance the ratio $R$ in unified models.
 The Higgs representation and bi-fermion of its conjugate representation are also listed.
 }
\end{table}

In Table 1, we list five possible bi-fermion couplings to the Higgs representations
for the anti-symmetric couplings to generate the off-diagonal elements in the left-handed squarks
in SU(5), flipped-SU(5) (whose gauge symmetry is SU$(5)\times$U(1)$_X$) and SO(10).
If the Higgs representation is not $\bf 45$ (for example, $\overline{\bf 10}$ for $(\overline{\bf 3},{\bf 2}, -\frac16)$),
the Yukawa coupling is not anti-symmetric,
and one should choose another gauge symmetry in the same row of the list.

As described, the right-handed FCNC should be small due to the constraints arising from the $B_s$-$\bar B_s$ mixing amplitudes.
Since each Higgs representation has a conjugate representation which can also have bi-fermion coupling,
we  list the corresponding conjugate bi-fermion couplings.
The conjugate bi-fermion couplings include the right-handed down-type quark $d^c$, barring
$({\bf 3},{\bf 3}, -\frac13) $ and
$(\overline{\bf 3},{\bf 2}, -\frac76)$.
In SU(5) or flipped-SU(5) model,
the conjugate bi-fermion couplings are ${\bf 10} \cdot \overline{\bf 5} \cdot \overline{\bf 45}$
and they are not necessarily unified to the anti-symmetric coupling, and the right-handed FCNC can be free in principle.
In SO(10), the conjugate coupling matrices are unified to the $\bf 120$ Higgs coupling,
and thus, the antisymmetric couplings are naively given by $({\bf 3},{\bf 3}, -\frac13) $ and
$(\overline{\bf 3},{\bf 2}, -\frac76)$ reps
in SO(10) model.
However, since the same reps. are included in $\overline{\bf 126}$ and ${\bf 126}$
(${\bf 45} \subset {\bf 126}$, $\overline{\bf 45} \subset \overline{\bf 126}$,
and ${\bf 45} + \overline{\bf 45} \subset {\bf 120}$)
and they can mix,
the linear combination of the light fields to generate the FCNC can be different between the conjugate and unconjugate reps.,
by adjusting the $\lambda {\bf 120}\cdot {\bf 126} \cdot {\bf 210}$ and $\bar \lambda {\bf 120}\cdot \overline{\bf 126} \cdot {\bf 210}$
couplings using $\lambda \gg \bar\lambda$.
Therefore, in general, all the five cases are possible in SO(10).

The $({\bf 3},{\bf 3}, -\frac13)+ c.c. $ can give rise to the proton decay operator $qqq\ell$,
and it may not be a good choice to make this rep. light to generate FCNC.
In flipped-SU(5),
$(\overline{\bf 3},{\bf 2}, -\frac16)+c.c.$
have the same quantum numbers as the would-be-Goldstone modes
to be eaten by the SU(5) gauge bosons.
Therefore, in the flipped-SU(5)-like vacua in SO(10), this rep. can be a good candidate to generate the FCNC.
We point out that $(\overline{\bf 6},{\bf 1}, -\frac13)$ and
$({\bf 8},{\bf 2}, \frac12)$ reps. are good candidates to increase the unification scale
and relax the bound due to the proton lifetime \cite{Dutta:2007ai}.

We note that the ${\bf 45}$ rep. in SU(5) and ${\bf 120}$ rep. in SO(10) contain the MSSM Higgs doublets
and the anti-symmetric coupling is a part of the linear combination of the Yukawa couplings
to generate the quark and lepton masses.
Since the mixings of the doublets are multiplied in the linear combination of the Yukawa couplings,
the original anti-symmetric couplings can be $O(1)$ and can provide large off-diagonal elements via RGE.

\subsection{Neutron and electron EDM}

%
The enhancement of the ratio $R$
can  impact the $t \to u \gamma(g)$ and $\tau \to e\gamma$ decays
rather than the $t \to c \gamma(g)$ and $\tau \to \mu\gamma$ decay processes.
However, both $t \to c\gamma (g)$ and $t \to u\gamma (g)$ branching fractions are tiny
due to the current bound on gluino and squark masses,
 and  it will be  hard to observe them.
%
The $\tau\to e\gamma$ process can be generated for
$({\bf 3},{\bf 3}, -\frac13)$ and $(\overline{\bf 3},{\bf 2}, -\frac76)$ couplings.
If the 13 off-diagonal element in the left-handed slepton mass matrix is turned on,
the chargino loop contribution can generate the branching fraction of $\tau\to e\gamma$
to be several times $10^{-9}$, by using the parameters from  Figs. 1 and 2.

The other impact of  13 generation mixings
is that they can generate  neutron and electron electric dipole moments (EDM).
The up and down quark EDM can be induced from the chargino diagrams
due to the 13 off-diagonal elements in left-handed squark mass matrix.
Since $V_{ub}$ or $V_{td}$ is multiplied to the amplitude,
the SUSY contribution is not very large ($\sim 10^{-28}$ $e\cdot{\rm cm}$), but
it is much larger than the SM predictions.
If there are 13 off-diagonal elements in both left- and right-handed squark mass matrices
due to the $({\bf 8},{\bf 2}, \frac12)$ FCNC source,
the amplitude can be much enhanced by a gluino diagram,
and it can make the neutron EDM  comparable to the current bound,
$|d_n| < 2.9 \times 10^{-26}$ $e\cdot{\rm cm}$ \cite{Baker:2006ts}.
The electron EDM can  also be enhanced by a neutralino diagram (for Bino components), if FCNC contributions arise from both
left- and right-handed charged-slepton mass matrices
induced by $(\overline{\bf 3},{\bf 2}, -\frac76)+c.c.$ couplings,
and it can be comparable to the current experimental bound $|d_e| < 8.7 \times 10^{-29}$ $e\cdot{\rm cm}$ \cite{Baron:2013eja}.

\section{Conclusion}

In conclusion, the anti-symmetric Yukawa interaction as a new source of FCNC
can explain the enhancement for the ratio of the branching fractions Br($B_d \to \mu^+\mu^-$)/Br($B_s \to \mu^+\mu^-$)
and the deviation of the experimental result from the SM prediction of the CP phase in the $B_d\to J/\psi K$ decay. The new interactions can be described by grand unified models, e.g., SO(10), SU(5), flipped SU(5) etc.
The enhancement of the ratio and natural realization of the $b\to d\gamma$ and $b\to s\gamma$ data
force the choice of the CP odd Higgs mass to be less than 1 TeV and $\tan\beta_H \alt 20$
for squark and gluino masses to be around 2 TeV.  The allowed parameter space satisfies   constraints arising from various hadronic and leptonic rare decay processes, $B$-$\bar{B}$,  $K$-$\bar{K}$ oscillations data and electric dipole moments of electron and neutron.

\section{Acknowledgement}

The work of B.D. is supported by DOE Grant DE-FG02-13ER42020. The work of Y.M. is supported by the Excellent Research Projects of
 National Taiwan University under grant number NTU-EPR-104R8915.

\end{document}